\documentclass[epj]{svjour}
\usepackage{graphics}
\begin{document}
\title{Cross section measurement of the $^{12}$C(p,$\gamma$)$^{13}$N reaction with activation in a wide energy range}
\author{Gy. Gy\"urky\inst{1} \and L. Csedreki\inst{1} \and T. Sz\"ucs\inst{1} \and G.G. Kiss\inst{1} \and Z. Hal\'asz\inst{1} \and Zs. F\"ul\"op\inst{1}
}                     
\offprints{}          
\institute{Institute for Nuclear Research (ATOMKI), H-4001 Debrecen, Hungary}
\date{Received: date / Revised version: date}
%
\abstract{
The CNO cycle is one of the fundamental processes of hydrogen burning in stars. The first reaction of the cycle is the radiative proton capture on $^{12}$C and the rate of this $^{12}$C(p,$\gamma$)$^{13}$N reaction is related to the $^{12}$C/$^{13}$C ratio observed e.g. in the Solar System. The low-energy cross section of this reaction was measured several times in the past, however, the experimental data are scarce in a wide energy range especially around the resonance at 1.7\,MeV. In the present work the $^{12}$C(p,$\gamma$)$^{13}$N cross section was measured between 300 and 1900 keV using the activation method. This method was only used several decades ago in the low-energy region. As the activation method provides the total cross section and has uncertainties different from those of the in-beam $\gamma$-spectroscopy technique, the present results provide a largely independent data set for future low-energy extrapolations and thus for astrophysical reaction rate calculations. 
} 
\maketitle
\section{Introduction}
\label{sec:intro}

In lack of heavier element, the first stars, produced from primordial material of the Big Bang, could burn their hydrogen fuel only through the so-called pp-chain reactions \cite{Iliadis2007}. In later stellar generations, however, the existence of chemical elements beyond helium allows other processes of hydrogen burning. The CNO cycle \cite{Wiescher2018,Wiescher2010} is the simplest of these processes which proceeds through the following nuclear reactions and radioactive decays:

\begin{enumerate}
	\item $^{12}$C(p,$\gamma$)$^{13}$N
	\item $^{13}$N($\beta^+\nu$)$^{13}$C
	\item $^{13}$C(p,$\gamma$)$^{14}$N
	\item $^{14}$N(p,$\gamma$)$^{15}$O
	\item $^{15}$O($\beta^+\nu$)$^{15}$N
	\item	$^{15}$N(p,$\alpha$)$^{12}$C
\end{enumerate}

This cycle is the dominant source of the released energy in main sequence stars heavier than about 1.5 solar masses, while in lower mass stars the pp-chains prevail. Nevertheless, even in the case of the Sun the CNO cycle is responsible for about 2\,\% of its energy output and solar models can also be tested by comparing the calculated CNO neutrino flux with observations which have become available recently \cite{Borexino2020}.

The rate of the CNO cycle is determined by its slowest reaction which is $^{14}$N(p,$\gamma$)$^{15}$O \cite{Gyurky2022}. On the other hand, good knowledge about the other participating reactions is also necessary. For example, the rate of the $^{12}$C(p,$\gamma$)$^{13}$N reaction is directly related to the $^{12}$C/$^{13}$C abundance ratio which comes about during a CNO burning episode. This ratio can be measured in the Solar System material \cite{Lodders2009} as well as in stellar and molecular cloud spectra \cite{Mikolaitis2012,Milam2005} and its value can be compared with astrophysical model calculations if the $^{12}$C(p,$\gamma$)$^{13}$N reaction rate is known. 

Depending on the astrophysical scenario, the typical temperatures for the CNO cycle are in the range of 10 - 100 GK. This translates into effective interaction energies between about 15 and 120\,keV for $^{12}$C(p,$\gamma$)$^{13}$N . This is the energy range where the reaction cross section must be known for the reaction rate determination. With the available experimental technique only the upper part of this energy region can be covered with measurements, the underground experiment of the LUNA collaboration will soon provide data down to about 80 keV \cite{Skowronski2022}. To lower energies, theory-based extrapolation is inevitable, which necessitates high precision cross section at higher energies. 

The $^{12}$C(p,$\gamma$)$^{13}$N reaction is dominated by two broad resonances at about 421 and 1556 keV center-of-mass energies\footnote{The higher energy resonance is actually the sum of two resonances corresponding to $^{13}$N excited stats at 3502 and 3547\,keV. Both states are, however, broad with total widths of 62 and 47\,keV, respectively, and hence they appear as one wide resonance in the experimental excitation function. In the following, the sum of these two resonances is meant when the ''high energy resonance'' is mentioned.}. The astrophysically relevant energy region lies on the low energy slope of the first resonance, where the extrapolation to lower energies needs to take into account the interference effect between the resonances and the direct capture component. Therefore, experimental information on the reaction cross section is needed all over the region of these resonances.

Not surprisingly, most of the previous astrophysics-motivated $^{12}$C(p,$\gamma$)$^{13}$N cross section measurements concentrated on the lowest measurable energies \cite{Bailey1950,Hall1950,Lamb1957,Hebbard1960,Vogl1963,Burtebaev2008}. Three experiments studied solely the higher energy resonance region \cite{Young1963,Poyarkov1987,Kiss1989}. The only available data set which covers the energy region of both resonances is from 1974 by Rolfs and Azuma \cite{Rolfs1974}. In order to satisfy the needs of high-precision astrophysical models, a new $^{12}$C(p,$\gamma$)$^{13}$N cross section measurement is thus required.

The aim of the present work was therefore to measure the $^{12}$C(p,$\gamma$)$^{13}$N cross section in a wide energy range covering both resonances and reaching the region of the low energy measurements. As it will be outlined below, the activation method was used in this work. Since Rolfs and Azuma used in-beam $\gamma$-spectroscopy, our results are largely independent from that data set and provide new experimental information for testing theoretical models.

In the next section the experimental procedure will be outlined. The obtained result will be presented in Sect.\,\ref{sec:results} and discussed in Sect.\,\ref{sec:discussion}.

\section{Experimental procedure}
\label{sec:experiment} 

\subsection{Activation method}
\label{sec:activation} 

The product of the $^{12}$C(p,$\gamma$)$^{13}$N reaction is radioactive, decays by positron emission to $^{13}$C with a half-life of 9.965\,$\pm$\,0.004\,min \cite{NDS_A13}\footnote{A new value of the $^{13}$N half-life was published very recently, after the conclusion of the present data analysis \cite{Long2022}. This new result is 9.951\,$\pm$\,0.003\,min, which differs by almost 3\,$\sigma$ from the previously adopted value, but this difference is not more than 0.14\,\%, which has no significant influence on the results of the present work.}. The decay is not followed by the emission of any $\gamma$-radiation, but the positron annihilation results in two 511\,keV $\gamma$-rays. The detection of these $\gamma$-rays makes it possible to measure the $^{12}$C(p,$\gamma$)$^{13}$N cross section by activation \cite{Gyurky2019}. 

The activation method has the advantage that it directly provides the total, angle integrated reaction cross section, which is the quantity needed for astrophysical reaction rate calculation. Moreover, this method is free from some systematic uncertainties encountered in in-beam $\gamma$-spectroscopy experiments. On the other hand, the activation technique does not provide information about partial cross sections of the various $\gamma$-transitions following the proton capture. 

For the $^{12}$C(p,$\gamma$)$^{13}$N reaction the activation method was used previously only at the lowest energies below $E_{\rm p}$\,=\,200\,keV \cite{Bailey1950,Hall1950}. In the present work this technique is applied to measure the cross section in a wide energy range, where this method has never been used before.

\subsection{Target preparation and characterization}
\label{sec:target} 

Solid state C targets were prepared by electron beam evaporation of graphite onto 0.5\,mm thick tantalum backings. Since natural carbon is dominated by the $^{12}$C isotope (98.93\,$\pm$\,0.08\,\% natural isotopic abundance \cite{Bohlke2001}), no isotopically enriched target material was necessary. Altogether five targets were prepared with similar thicknesses. 

For the absolute cross section measurement the number density of the target atoms, i.e. the target thickness, must be known precisely. As the natural targets contain 1.07\,\% $^{13}$C, target thicknesses were measured with Nuclear Resonant Reaction Analysis (NRRA) utilizing the narrow resonance in the $^{13}$C(p,$\gamma$)$^{14}$N reaction at $E_{\rm p}$ = (1747.6 $\pm$ 0.9)\,keV. For the NRRA measurement the Tandetron accelerator of Atomki was used which provided the proton beam at energies around the resonance with a few $\mu$A intensity. More details about the method can be found in \cite{Ciani2020}, where a similar target preparation and characterization experiment is presented in relation to a $^{13}$C($\alpha$,n)$^{16}$O cross section measurement.

%
\begin{figure}
\resizebox{0.95\columnwidth}{!}{\includegraphics{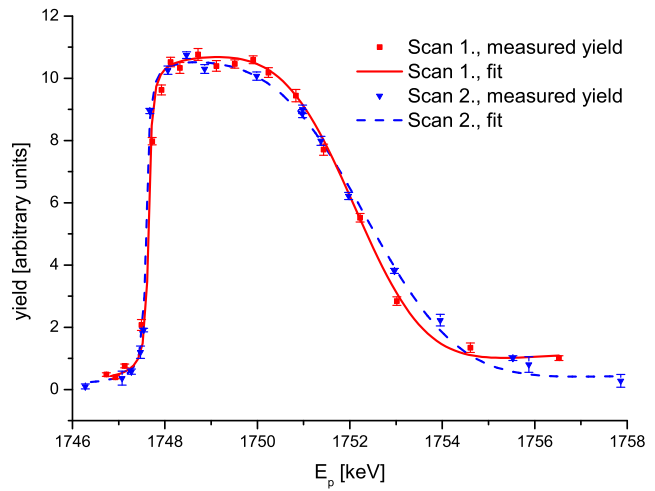}}
\caption{Profile of the $E_{\rm p}$ = (1747.6 $\pm$ 0.9)\,keV resonance measured on an evaporated C target before (red squares and solid line) and after (blue triangles and dashed line) the cross section measurement campaign. See text for details.}
\label{fig:resonance}       
\end{figure}

Figure\,\ref{fig:resonance} shows the result of the NRRA measurements on one of the targets. 
Besides the measured yields, fit functions are also shown which were used to determine the target thickness based on the width of the resonance profiles. For the thickness calculation a pure carbon layer was assumed and the proton stopping power in carbon was taken from the SRIM code \cite{SRIM}. The stopping power provided by SRIM is in perfect agreement with the precise experimental data of S. Gorodetzky \textit{et al.} \cite{Gorodetzky1967} which have at most 2.3\,\% uncertainty in the energy range of the resonance. Therefore, this 2.3\,\% was taken as the stopping power uncertainty.

Similar profiles as presented in Fig.\,\ref{fig:resonance} were measured for all five targets and target thickness values between 1.03$\cdot$10$^{18}$ and 1.52$\cdot$10$^{18}$ atoms/cm$^2$ were obtained. Taking into account the uncertainty of the stopping power and that of the resonance profile fit, the target thicknesses were determined with 5\,\% precision.

Ion beam bombardment of a solid state target may cause target degradation, as it was clearly demonstrated by the $^{13}$C($\alpha$,n)$^{16}$O experiment of the LUNA collaboration \cite{Ciani2020}. In order to investigate the target stability, the resonance profiles were measured not only on the freshly prepared targets, but also after the cross section measurement campaign. The results of the two measurements, as shown in Fig.\,\ref{fig:resonance} with labels Scan 1 and Scan 2, respectively, indicated that no observable target degradation occurred. This is in line with the findings of \cite{Ciani2020}. In the present case the maximum collected proton beam charge on one target was about 0.2\,C, while in \cite{Ciani2020} observable target degradation was found roughly after collecting charge of the order of 1\,C. Moreover, proton beams used in this case cause much less deterioration than alpha beams used in \cite{Ciani2020}. Therefore, the stability of our targets was expected and also experimentally proved.

\subsection{Irradiations}
\label{sec:irrad} 

The proton beam for the $^{12}$C(p,$\gamma$)$^{13}$N cross section measurements was provided by the Tandetron accelerator of Atomki \cite{Rajta2018,Biri2021}. The beam energy calibration of the accelerator has recently been done to a precision of better than 1\,keV \cite{Csedreki2020}. The energy spread of the beam is about 160\,eV, based on the yield curve measurements of narrow nuclear resonances \cite{Rajta2018}. The studied energy range between 300 and 1900\,keV was covered in 50-100\,keV energy steps. Depending on the cross section to be measured, the beam intensity varied between 1 and 5\,$\mu$A. 

\begin{figure}
\resizebox{\columnwidth}{!}{\includegraphics{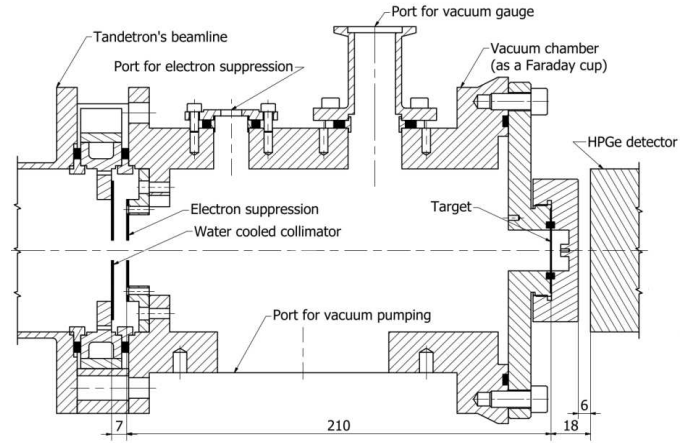}}
\caption{Drawing of the target chamber with the position of the HPGe detector behind the target. The indicated dimensions are in millimeters.}
\label{fig:chamber}       
\end{figure}

The targets were placed into the activation chamber which served as a Faraday cup. The drawing of the setup is shown in Fig.\,\ref{fig:chamber}. The beam enters the chamber through a water-cooled collimator of 5\,mm in diameter which guarantees that the beam hits only the target. The number of protons impinging on the target was thus measured by integrating the charge entering the chamber with an ORTEC model 439 digital current integrator. 

As the half-life of $^{13}$N is relatively short, a cyclic activation method, similar to the one presented in Refs. \cite{Gyurky2022,Gyurky2019b} was used. Using an automatic beam shutter, the proton beam bombarded the target for 20 minutes which was followed by a 40 min beam-off period used for the decay counting. At a given proton energy, this 60 min cycle was repeated until sufficient counting statistics was achieved (the maximum number of cycles was 17 at the lowest measured cross section). During the whole activation, the beam charge was integrated and recorded in every 10 seconds in order to take into account any beam intensity fluctuation in the activation analysis.

In the studied energy range, the highest cross section is found at the top of the first resonance at $E_{\rm p}$\,=\,456\,keV. In order to exploit the high yield near this energy, all five targets were irradiated with a proton beam of $E_{\rm p}$\,=\,468.5\,keV. The cross sections obtained from these runs (see Sect. \ref{sec:results}) provide an independent check on the reliability of the measured cross sections using different targets. 

\subsection{Detection of the $^{13}$N decay radiation}
\label{sec:detection} 

The 40 min beam-off periods were used to measure the decay of the $^{13}$N reaction product. An HPGe detector with 100\,\% relative efficiency was used for this purpose, which detected the 511\,keV $\gamma$-radiation created by the annihilation of positrons from the $^{13}$N decay. The cyclic activation was carried out without removing the target from the activation chamber. The detector was placed in close geometry just outside the vacuum chamber behind the target along the beam direction. The front face of the detector was 18\,mm far from the target, see Fig.\,\ref{fig:chamber}.  

\begin{figure}
\resizebox{0.95\columnwidth}{!}{\includegraphics{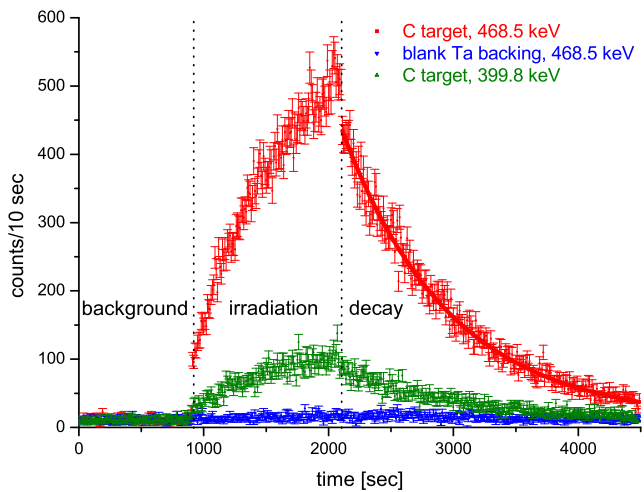}}
\caption{Number of detected 511\,keV $\gamma$-rays as a function of time in the case of a C target (red) and a blank Ta backing (blue) at $E_{\rm p}$\,=\,468.5\,keV bombarding energy as well as a C target at $E_{\rm p}$\,=\,399.8\,keV bombarding energy (green). In the first case the fitted exponential decay curve using the literature half-life of $^{13}$N is also shown as a red solid line.}
\label{fig:decay}       
\end{figure}

The number of events detected in the 511\,keV positron annihilation peak was recorded as a function of time with a 10 sec time basis. In the case of one activation cycle, such recorded yield functions can be seen for two different proton energies in Fig.\,\ref{fig:decay}. Three regions can be distinguished in the figure. Before the start of the irradiation the detector records the 511\,keV $\gamma$-rays from environmental radiation, which is used as the background component in the analysis. During the irradiation phase the increase of the 511\,keV yield can be seen as a consequence of the accumulated $^{13}$N activity. In this period, beam-induced background also contributes to the detected events at 511\,keV. This is caused mainly by positron annihilation resulting from pair production of prompt high-energy $\gamma$-rays from nuclear reactions on target impurities. This contribution can be seen in Fig\,\ref{fig:decay} by the mismatch of the counting rates between the background, irradiation and decay periods. Therefore, in order to avoid the uncertainty associated with beam-induced background, only the decay periods were used for the analysis.

In the beam-off period the decay of $^{13}$N can be followed. This part - and in the case of multiple cycles, several other decay parts - were used to calculate the number of created $^{13}$N isotopes which is needed for the cross section determination. Based on the known half-life of $^{13}$N, the decay parts were fitted with an exponential function superimposed on a constant background. Such a fitted curve can be seen in Fig\,\ref{fig:decay} in the case of the 468.5\,keV irradiation (red line). No deviation form the expected $^{13}$N decay curve was observed indicating that no other positron emitter isotopes were produced during the beam bombardment in any significant amount, i.e. such a contribution was found to be well below 1\,\%. 

\begin{figure}
\resizebox{0.9\columnwidth}{!}{\includegraphics{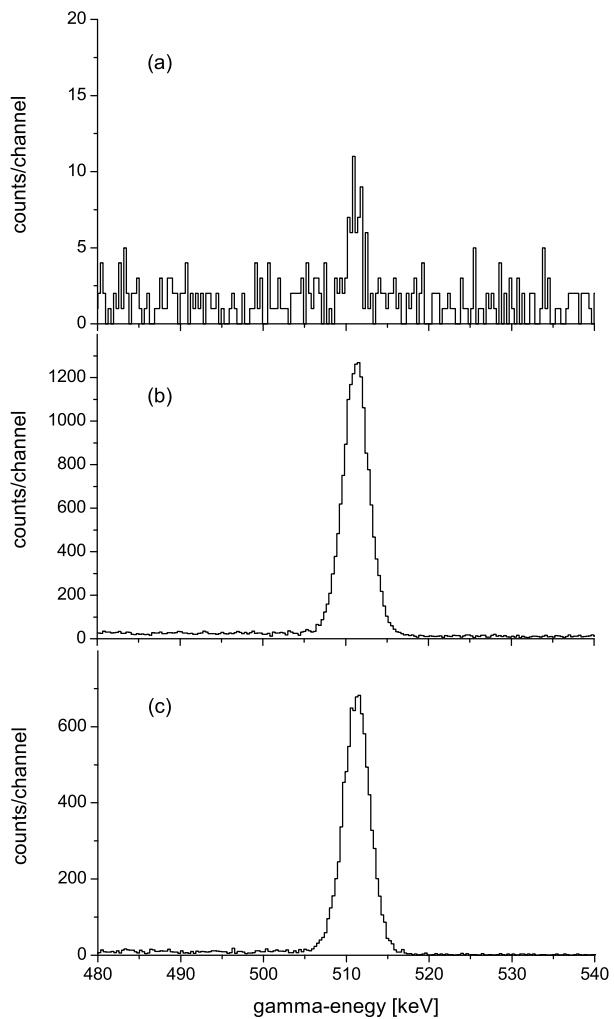}}
\caption{Relevant part of the $\gamma$-spectra around 511\,keV measured for a counting time of five minutes in three cases: (a): laboratory background, (b): spectrum recorded during the irradiation with $E_{\rm p}$\,=\,468.5\,keV protons started 10 minutes after the beginning of the irradiation, (c): spectrum recorded after this irradiation during the $^{13}$N decay, started 10 minutes after the stop of the irradiation. Note the different vertical scales of the three spectra. One channel corresponds to 0.31\,keV.}
\label{fig:spectra}       
\end{figure}

As an example, Fig.\,\ref{fig:spectra} shows the relevant part of the measured $\gamma$-spectra around 511\,keV in three cases: Laboratory background (panel a), spectrum recorded during the irradiation with $E_{\rm p}$\,=\,468.5\,keV protons (panel b) and after this irradiation during the $^{13}$N decay (panel c). All three spectra were taken for 5 minutes. The number of events recorded in the 10 second intervals described above was determined by putting a gate at the region of the 511\,keV peak. 

As carbon is a common contaminant in vacuum systems, for example on the surface of collimators bombarded by the beam, spurious events from the $^{12}$C(p,$\gamma$)$^{13}$N reaction not originating from the target itself must be identified. For this purpose, a background measurement at $E_{\rm p}$\,=\,468.5\,keV was carried out using a blank Ta backing in place of the target. The result of this measurement is also shown in Fig.\,\ref{fig:decay}. As it can clearly be seen, the number of 511\,keV $\gamma$-ray events remained at the level of the laboratory background during the whole run, indicating that the contribution of spurious events to the measured reaction yield remains well below 1\,\%. 

The detection efficiency of the $\gamma$-detector registering the 511\,keV $\gamma$-rays is a crucial quantity for the cross section determination. As it is discussed in Ref. \cite{Gyurky2019}, the positron annihilations resulting in the 511\,keV $\gamma$-rays do not occur in a point-like geometry. Positron may be emitted by the decaying $^{13}$N nuclei towards the vacuum chamber and the annihilation takes place at various positions on the surface of the chamber. Therefore, the usual approach of measuring the efficiency with calibrated radioactive sources is not applicable in the case of the present complicated geometry. An indirect method similar to the one presented in \cite{Gyurky2019b} was thus used for the efficiency measurement.

A target was irradiated at $E_{\rm p}$\,=\,468.5\,keV, close to the top of the first resonance in order to maximize the collected $^{13}$N activity. The decay was first measured for 20 minutes (about two half-lives) in the counting geometry applied in all other measurements, i.e. keeping the target in the vacuum chamber. The target was then removed from the chamber and placed in front of the same detector, but in this case covered with a 1\,mm thick Ta sheet. From the Q-value of the $^{13}$N decay \cite{NDS_A13}, the maximum energy of the positrons is about 1.2\,MeV while the average is 492\,keV. The maximum range of positrons in tantalum with such energies is less than 400\,$\mu$m \cite{Tabata1972,Saxena2010}. Thus, the 1\,mm thick Ta sheet as well as the 0.5\,mm thick Ta target backing stops all the emitted positrons and hence the 511\,keV $\gamma$-ray emission took place in a quasi point-like geometry with less than 1\,mm extension. This size is comparable with the typical 2\,mm of diameter active area of calibration sources. In the same geometry the efficiency of the detector was measured with calibrated sources and based on this procedure the efficiency in the counting geometry could also be determined. The absolute efficiencies in these two geometries were measured to be 11.4\,\% and 2.81\,\%, respectively. The efficiency measurement was carried out at the end of the experimental campaign, the detector was kept in a fixed position during the cross section measurements and moved only for the efficiency determination. Therefore, no uncertainty due to the detector repositioning was introduced. The efficiency measured this way has an uncertainty of 5\,\%, a value which takes into account the statistical uncertainty of decay counting both inside and outside the chamber as well as the uncertainty of the calibration source activities and the related uncertainty of the detector efficiency in the latter geometry.

\section{Experimental results}
\label{sec:results} 

The $^{12}$C(p,$\gamma$)$^{13}$N cross section was measured in the proton energy range between 300 and 1900\,keV covering the region of the two resonances and reaching the energies of some of the previous low-energy experiments \cite{Hebbard1960,Vogl1963,Burtebaev2008} and the upcoming LUNA data \cite{Skowronski2022}. Table \ref{tab:results} shows the result. The first column contains the proton beam energy while in the second one the energy loss of the beam in the carbon target layer can be found. The third column lists the effective proton energies, which are discussed in Sec.\,\ref{sec:discussion}. In the last three columns the obtained cross sections, their statistical and total uncertainties are given.

\begin{table}
\caption{The measured cross section of the $^{12}$C(p,$\gamma$)$^{13}$N reaction. See text for further details.}
\label{tab:results}
\begin{tabular}{llllll}
\hline
$E_{\rm p}$ \hspace{2mm} & $\Delta E$ \hspace{2mm} & $E_{\rm eff.}$  & \hspace{2mm}  $\sigma$ \hspace{2mm} & $\Delta\sigma^{\rm stat.}$ \hspace{2mm} & $\Delta\sigma^{\rm total}$ \\
\cline{4-6}
\hspace{0mm}[keV] & [keV] & [keV] & \multicolumn{3}{c}{$\mu$barn} \\
\hline
300.0	&	11.8	&	294.6	&	0.380	&	0.014	&	0.032	\\
350.0	&	10.4	&	345.2	&	1.102	&	0.027	&	0.089	\\
399.8	&	10.0	&	395.3	&	5.126	&	0.041	&	0.396	\\
449.8	&	11.5	&	445.1	&	57.60	&	0.30	&	4.43	\\
468.5	&	9.7	&	463.4	&	94.67	&	0.29	&	7.28	\\
468.5	&	7.6	&	464.5	&	92.25	&	0.54	&	7.11	\\
468.5	&	9.0	&	463.8	&	95.62	&	0.42	&	7.36	\\
468.5	&	11.2	&	462.6	&	90.56	&	0.68	&	6.99	\\
468.5	&	8.7	&	464.0	&	95.95	&	0.82	&	7.42	\\
500.0	&	9.3	&	494.9	&	25.00	&	0.12	&	1.92	\\
549.9	&	7.8	&	545.8	&	5.963	&	0.070	&	0.463	\\
590.0	&	8.3	&	585.8	&	3.378	&	0.022	&	0.260	\\
650.0	&	7.8	&	646.0	&	1.855	&	0.018	&	0.144	\\
700.0	&	5.8	&	697.1	&	1.225	&	0.037	&	0.101	\\
750.0	&	6.6	&	746.7	&	0.911	&	0.015	&	0.072	\\
800.0	&	5.4	&	797.3	&	0.722	&	0.022	&	0.060	\\
850.0	&	6.1	&	846.9	&	0.633	&	0.016	&	0.051	\\
899.9	&	5.7	&	897.0	&	0.514	&	0.022	&	0.045	\\
900.0	&	6.4	&	896.8	&	0.550	&	0.016	&	0.045	\\
999.9	&	5.4	&	997.2	&	0.421	&	0.015	&	0.035	\\
1000.0	&	4.7	&	997.6	&	0.467	&	0.018	&	0.040	\\
1099.9	&	4.5	&	1097.7	&	0.417	&	0.019	&	0.038	\\
1099.9	&	6.7	&	1096.6	&	0.442	&	0.013	&	0.036	\\
1200.0	&	6.3	&	1196.8	&	0.444	&	0.018	&	0.039	\\
1300.0	&	6.0	&	1297.0	&	0.555	&	0.017	&	0.046	\\
1400.0	&	4.4	&	1397.8	&	0.797	&	0.036	&	0.071	\\
1500.0	&	3.6	&	1498.1	&	1.567	&	0.055	&	0.132	\\
1600.0	&	3.9	&	1598.1	&	4.156	&	0.088	&	0.331	\\
1650.0	&	4.8	&	1647.7	&	12.43	&	0.10	&	0.96	\\
1700.0	&	3.6	&	1698.2	&	35.59	&	0.42	&	2.77	\\
1750.0	&	4.6	&	1747.6	&	7.262	&	0.079	&	0.563	\\
1800.0	&	4.5	&	1797.7	&	2.985	&	0.097	&	0.249	\\
1850.0	&	4.4	&	1847.8	&	1.847	&	0.041	&	0.148	\\
1899.8	&	3.4	&	1898.0	&	1.535	&	0.040	&	0.125	\\

\hline
\end{tabular}
\end{table}

As discussed in Sect.\,\ref{sec:irrad}, at $E_{\rm p}$\,=\,468.5\,keV energy all five targets were used for cross section measurements. These values are listed separately in the table. At three energies (900, 1000 and 1100\,keV) near the cross section minimum between the two resonances, two measurements were carried out. These results are also listed separately. 

An uncertainty of 1\,keV can be assigned to the primary proton beam energies listed in the first column, based on the recent accelerator calibration \cite{Csedreki2020}. The energy loss in the target shown in the second column was calculated from the measured target thicknesses and the stopping power taken from the SRIM code. From these two quantities the uncertainty of the energy loss amounts to 6\,\%, i.e. between 0.2 and 0.7\,keV. In the last two columns the statistical and total cross section uncertainties are given. The statistical uncertainty stems from the decay counting and for the total uncertainty the following components are added quadratically: target thickness (5\,\%), detection efficiency (5\,\%) and beam current integration (3\,\%). Other sources of uncertainty (like the $^{13}$N decay parameters or the $^{12}$C abundance ratio in natural carbon) are well below 1\,\% and therefore neglected.

\begin{figure*}
\resizebox{0.95\textwidth}{!}{\includegraphics{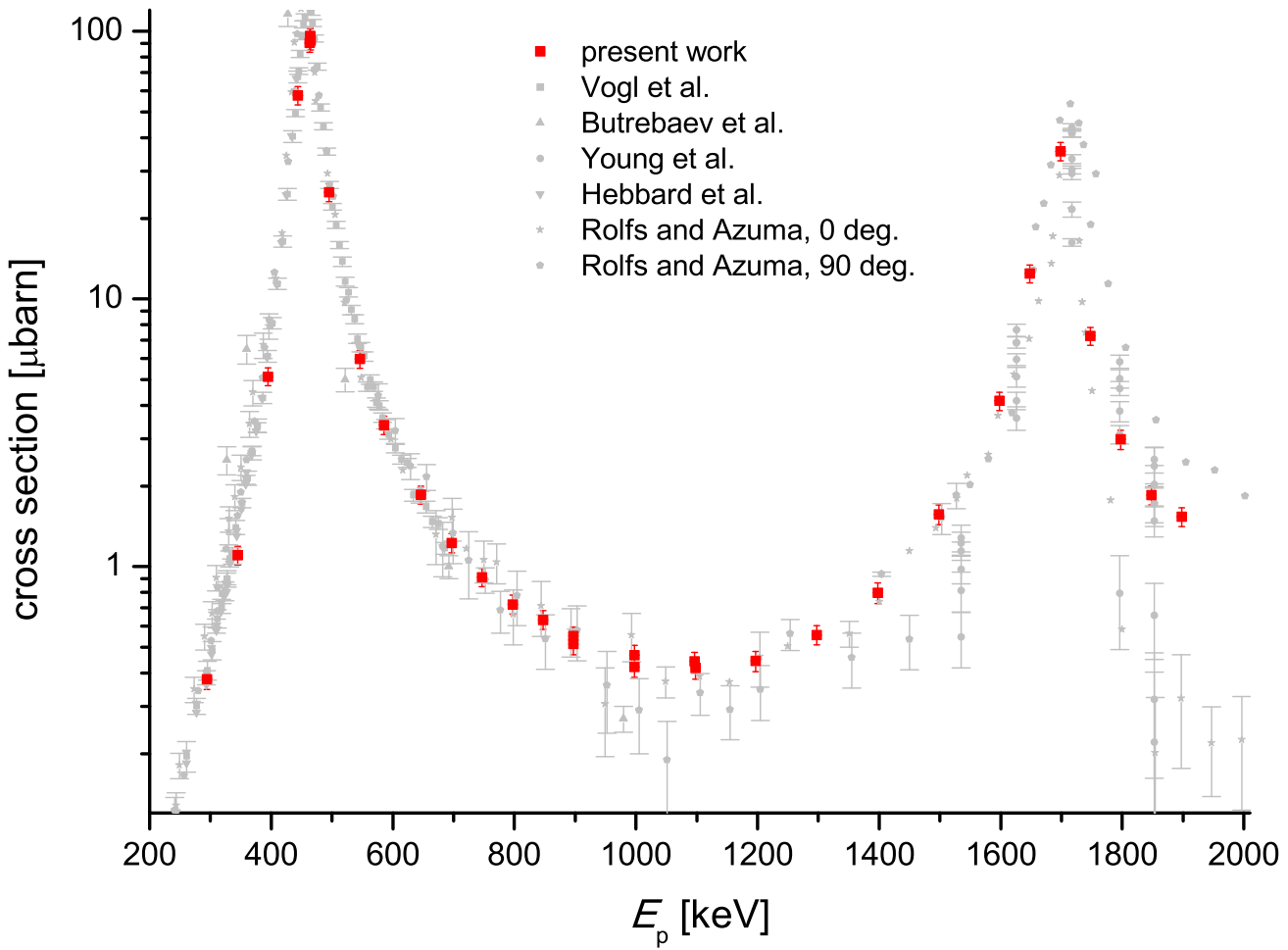}}
\caption{Total cross section of the $^{12}$C(p,$\gamma$)$^{13}$N reaction measured in this work and taken from the literature. In the case of the present results, the x coordinate of the plotted points correspond to the effective proton energies listed in the third column of Table \ref{tab:results} and discussed in Sec.\,\ref{sec:discussion}. See text for further details. }
\label{fig:results}       
\end{figure*}

The results obtained in the present work along with the available literature data in the studied energy range are presented in Fig.\,\ref{fig:results}. The uncertainty of the energy is smaller than the size of the symbols. In order to increase visibility, the present data are plotted in red, while all other data are shown in light gray using different symbols. The numeric values for the literature data are taken from the exfor database \cite{exfor}. In the case of Refs.\,\cite{Hebbard1960,Young1963,Rolfs1974} the exfor data are obtained from figure digitization which involves a certain uncertainty. In those cases where the experimental errors were smaller than the symbols in the figures, no uncertainties are given in exfor and therefore not plotted in Fig.\,\ref{fig:results}. 


The activation method used in the present work provides the total, angle integrated cross section and this is plotted in Fig.\,\ref{fig:results}. In the works of Refs.\,\cite{Young1963,Rolfs1974} differential cross sections measured at various detection angles are presented and these values are compiled in exfor. In order to compare those data with the total cross sections of the present work and with those of Refs.\,\cite{Hebbard1960,Vogl1963,Burtebaev2008}, the differential cross sections are multiplied by 4$\pi$. Since Rolfs and Azuma \cite{Rolfs1974} measured only at two angles (0 and 90 degrees), they are shown separately in the figure, while the data of Young \textit{et al.} \cite{Young1963} measured at several angles are shown as one data set with a single symbol.

In general, there is a good agreement between the present results and the literature data. In some energy regions, especially between the two resonances and above the second resonance, the available literature data show significant differences and are available only with large error bars. The present, high precision results provide a good data set to constrain the theoretical calculations at these energies.

\section{Discussion and conclusions}
\label{sec:discussion}

The presence of the two resonances in the studied energy range means that the cross section is a steeply varying function of the energy. Even though the energy loss of the beam in the targets is relatively low, below 12 keV (see the second column of table \ref{tab:results}), the cross section changes significantly within such an energy interval, especially close to the top of the resonances. Therefore, the determination of the effective interaction energy \cite{Lemut2008} is not trivial. A simplified R-matrix calculation was carried out aiming mainly at the effective energy determination, as such a calculation gives the energy dependence of the cross section in the energy regions covered by the targets. The effective proton energies listed in the third column of table \ref{tab:results} were thus obtained using the definition of median energy in Ref.\,\cite{Lemut2008} based on the R-matrix calculation outlined in the next paragraphs.

The R-matrix calculation was carried out with the AZURE2 computer code \cite{azure}. Altogether six energy levels of $^{13}$N were considered. Besides the ground state, the three levels corresponding to the resonances in the energy range of the present work were taken into account at 2364.9, 3502 and 3547\,keV energies, with spin and parity values of 1/2$^+$, 3/2$^-$ and 5/2$^+$, respectively. In addition, two background poles with 1/2$^+$ and 3/2$^+$ spin and parity were considered and placed at 8\,MeV to account for the contributions of the higher energy 1/2$^+$ and 3/2$^+$ broad levels. Other narrower exited states in the same energy region were omitted as their contribution to the studied energy range is negligible.

\begin{figure}
\resizebox{0.85\columnwidth}{!}{\includegraphics{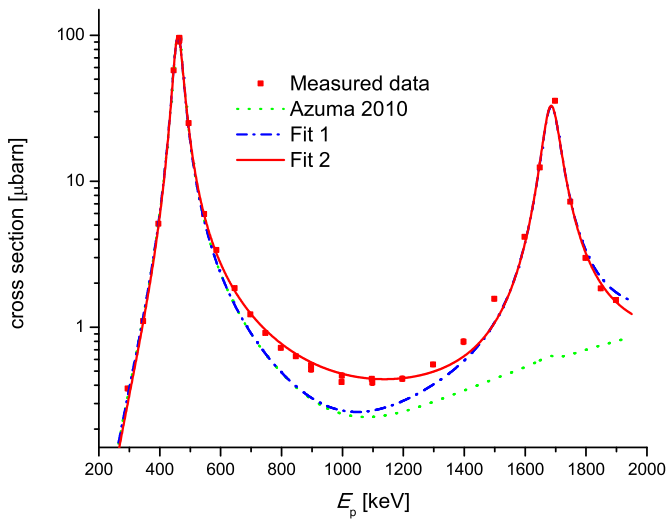}}
\caption{R-matrix calculations of the $^{12}$C(p,$\gamma$)$^{13}$N cross section along with the present experimental data. See text for details.}
\label{fig:Rmatrix}       
\end{figure}

Using the parameters presented in Ref. \cite{azure}, an initial excitation function was produced. Since the capture data above $E_{\rm p}$\,=\,600\,keV was not considered in that parameter set, this curve does not reproduce the higher energy resonance at $E_{\rm p}$\,=\,1.7\,MeV at all. The curve labeled Azuma 2010 shows this calculation in Fig. \ref{fig:Rmatrix} along with the experimental data of the present work. As a second step, a fit of the present data was performed where the Azuma 2010 parameters were kept fixed and only the gamma-widths of the 3502 and 3547\,keV levels were varied. It is clearly seen that without attributing strength to the higher energy background poles, the fit underestimates the region between the two resonances (Fit 1 in Fig.\,\ref{fig:Rmatrix}).  Nevertheless, this fit was used to obtain the effective energies of the cross section data points measured in this work. These values are listed in the third column of table \ref{tab:results} and used for plotting the experimental results in Figs.\,\ref{fig:results} and \ref{fig:Rmatrix}.  

A second fit was performed using the data with the updated $E_{\rm eff.}$. In this case the strengths (proton- and gamma-widths) of the two background poles were also varied. The result of this last fit is presented in Fig. \ref{fig:Rmatrix} as Fit 2. The effective energies were recalculated also with this second fit, but the change was in the order of 1 eV, thus completely negligible compared to the energy uncertainty.

The fit describes the experimental data quite well. The present data points alone do not place a strong constraint on the level parameters of the resonances in the measured energy range. When those are left free for the fit, none of them changes significantly compared to the uncertainty attributed to them in Ref.\,\cite{azure}. As noted by Azuma \textit{et al.} \cite{azure}, there is a discrepancy in the literature for the 3/2$^-$ resonance energy \cite{Kiss1989,Rolfs1974}. There is a tension between the resonance position derived from the scattering data \cite{Meyer1976}, and by Rolfs and Azuma radiative capture data \cite{Rolfs1974}. Our new data set is compatible with the Azuma \textit{et al.} \cite{azure} resonance position for the 3/2$^-$ level, thus compatible with the scattering data set \cite{Meyer1976}, and the Kiss \textit{et al.} data \cite{Kiss1989}, pointing to the direction that the Rolfs and Azuma data set may have some problems is the energy scale. A comprehensive R-matrix fit would require more data near the top of the resonances to constrain their position and strength. Such a measurement concentrating on the resonance position is in progress in Atomki using in-beam $\gamma$-spectroscopy and will be presented in a forthcoming publication. 

\section{Summary}

In the present work the cross section of the $^{12}$C(p,$\gamma$)$^{13}$N reaction was measured in the energy range between $E_{\rm p}$\,=\,300 and 1900\,keV using the activation method. This method -- which has never been used in the studied energy range -- provides directly the total cross section and the obtained results are substantially independent from those of the in-beam $\gamma$-spectroscopy experiments. 

A comprehensive R-matrix fit would require more data near the top of the studied resonances to constrain their parameters. Such a measurement concentrating on the resonance position and width is in progress in Atomki using in-beam spectroscopy and will be presented in a forthcoming publication. Owing to its good overall precision, the present data set can be used to constrain the theoretical calculations based on the forthcoming in-beam $\gamma$-spectroscopy data from three laboratories\footnote{Besides the low energy experiment of the LUNA collaboration and the resonance position study at Atomki quoted above, the $^{12}$C(p,$\gamma$)$^{13}$N reaction is being studied with in-beam $\gamma$-spectroscopy at the Felsenkeller laboratory in Dresden, Germany.}. 

\section*{Acknowledgments}

This work was supported by the National Research, Development and Innovation Office NKFIH (contract numbers K134197 and PD129060) and by the New National Excellence Program of the Ministry of Human Capacities of Hungary
(\'UNKP-22-5-DE-428). T.S. acknowledges support from the Bolyai research fellowship. The financial support of the Hungarian Academy of Sciences (Infrastructure grants), and the Economic Development and Innovation Operational Programme (GINOP-2.3.3–15-2016-00005) grant, co-funded by the EU, is also acknowledged.

\end{document}